\documentclass[%
reprint,
superscriptaddress,
amsmath,amssymb,
aps,
pra,
]{revtex4-1}
\usepackage{graphicx}% Include figure files
\usepackage{dcolumn}% Align table columns on decimal point
\usepackage{bm}% bold math
\usepackage[colorlinks=true,linkcolor=red,citecolor=blue]{hyperref}% add hypertext capabilities
\usepackage[mathlines]{lineno}% Enable numbering of text and display math
\usepackage{booktabs}

\usepackage{bbm}
\usepackage{dsfont}
\usepackage{amsmath}
\usepackage{verbatim}
\usepackage{graphicx}

\usepackage{braket,amsmath}
\usepackage[dvipsnames, svgnames, x11names]{xcolor}
\usepackage{array}

\usepackage{mathrsfs}
\usepackage{amsmath}
\usepackage{graphicx}% Include figure files
\usepackage{dcolumn}% Align table columns on decimal point
\usepackage{bm}% bold math
\usepackage[capitalise]{cleveref}
\crefname{section}{Sec.}{Secs.}
\crefname{appendix}{App.}{Apps.}

\begin{document}

\title{Engineering near-unitary one-axis twisting evolution via a driven Tavis-Cummings model}

\author{Jinfeng Liu}
\affiliation{Department of Physics, Wenzhou University, Zhejiang 325035, China}

\author{Yan Mu}
\affiliation{Department of Physics, Wenzhou University, Zhejiang 325035, China}

\author{Lili Song}
\affiliation{Department of Physics, Wenzhou University, Zhejiang 325035, China}

\author{Gang Liu}
\affiliation{School of physics, Zhejiang University, Hangzhou 310027, China}

\author{Mingfeng Wang}
\altaffiliation{mfwang@wzu.edu.cn}
\affiliation{Department of Physics, Wenzhou University, Zhejiang 325035, China}

\date{\today }

\begin{abstract}
One-axis twisting (OAT) interaction is a pivotal resource for manipulating quantum states of atomic ensembles, enabling spin squeezing, atomic-cat-state generation, and weak-phase amplification. Current implementations of OAT dynamics predominantly rely on the Tavis-Cummings model of light-atoms coupling; however, this approach inevitably introduces an additional Stark term that entangles the light with the atoms, which compromises the unitarity of OAT evolution and thereby degrades the OAT-based control precision. Here we propose a scheme based on a driven Tavis-Cummings model to achieve near-unitary OAT evolution. We demonstrate that both constant and time-varying driving of an atoms-cavity hybrid system can realize near-unitary OAT evolution, albeit with distinct coupling strength. Furthermore, when atomic dissipation is taken into account, we find that the time-varying-driving scheme exhibits superior resistance to decoherence. Our approach is broadly applicable to a variety of atomic platforms, including cold atoms, trapped ions, and nitrogen-vacancy centers.
\begin{description}
	\item[PACS numbers]
	42.50.Pq, 42.50.Dv, 03.67.Bg
\end{description}
\end{abstract}

\pacs{Valid PACS appear here}

\maketitle

\section{Introduction}
Entangled states of atomic ensembles have important applications in quantum information processing  \cite{RevModPhys.77.513}, quantum sensing \cite{n0,n1,NP1}, and quantum simulation \cite{pezze2018quantum}. Therefore, developing efficient methods to prepare such states has become a central research focus. To date, several methods have been developed to achieve this goal, including quantum non-demolition (QND) measurements \cite{grangier1998quantum,kono2018quantum,appel2009mesoscopic,schleier2010states,chen2011conditional}, transferring entangled states of light to atomic ensembles~\cite{ficek2002entangled,sherson2007deterministic,hald1999spin}, and evolution via spin-spin nonlinear interactions~\cite{sanz2003entanglement,liu2019spin}. Among them, the last approach is attracting particular attention due to its ability to generate highly entangled atomic states and its feasibility to implement in various atomic systems \cite{leroux2010implementation,braverman2019near,n2}. Although a variety of nonlinear interactions exist, the OAT interaction is the most extensively studied, as it enables the generation of spin squeezed state (SSS) \cite{pezze2018quantum,kitagawa1993squeezed,opatrny2015twisting,opatrny2015spin,wu2015spin,PhysRevA.50.67,PhysRevA.46.R6797,liu2011spin}, the creation of atomic Greenberger-Horne-Zeilinge (GHZ) state \cite{greenberger1990bell,PhysRevA.56.2249}, and the  amplification of weak phase \cite{PhysRevLett.116.053601,n1}. Several approaches have been proposed to realize OAT dynamics in atomic ensembles. These include, for example, utilizing mechanisms such as atom-atom collisions \cite{n1,n4}, quantum erasure \cite{leroux2012unitary,PhysRevA.96.013823}, and Tavis-Cummings (TC) atom-light interactions \cite{dong2022dynamics,genes2003spin,van2018microwave,wang2020controllable,fink2009dressed,groszkowski2020heisenberg,marinkovic2023singly,PhysRevLett.110.156402,lewis2018robust,koppenhofer2023revisiting}.

The standard TC model describes a collection of identical atoms interacting equally with a single bosonic mode via a Jaynes-Cummings interaction \cite{PhysRev.170.379}. Due to the relative ease implementation of the Jaynes-Cummings interaction in a wide variety of atomic platforms, the TC model has emerged as a primary method for engineering OAT dynamics  \cite{PhysRevLett.110.156402,lewis2018robust,dooley2016hybrid}. The basic principle of OAT engineering with the TC model can be understood qualitatively as follows: the bosonic mode plays the role of a quantum bus, which imprints atomic information onto itself and thereby mediates nonlinear evolution of the atoms.
However, during this process, it is inevitable that the bosonic mode get entangled with the atoms, which introduces an extra Stark term (originating from ac Stark shift) alongside the desired OAT interaction. Although it is well-known that the Stark term can induce atomic dephasing, its precise impact on the creation of entangled states has not yet been thoroughly explored.

In this paper, we revisit the TC approach to OAT engineering. Our work complements previous studies both by thoroughly investigating the influence of the Stark term on the creation of different types of atomic entangled states, as well as proposing an alternative method for implementing near-unitary OAT evolution. In terms of Stark term's influences, we investigate how the initial state of the bosonic mode affects the generation of entangled states. We show that any initial state other than the Fock states will break the unitarity of the OAT evolution. The nonunitary nature imposes a fundamental limit on spin squeezing, leading to a spin-squeezing scaling that is significantly worse than that achievable under unitary OAT. When it comes to the GHZ-state generation, the nonunitary nature has an even more pronounced negative effect, causing the fidelity of state preparation to decay exponentially with the amplitude of the input coherent state. Therefore, it is vital to suppress or eliminate the influence of the Stark term.
\begin{figure}[t]
	\centering
	\includegraphics[width=0.98\linewidth]{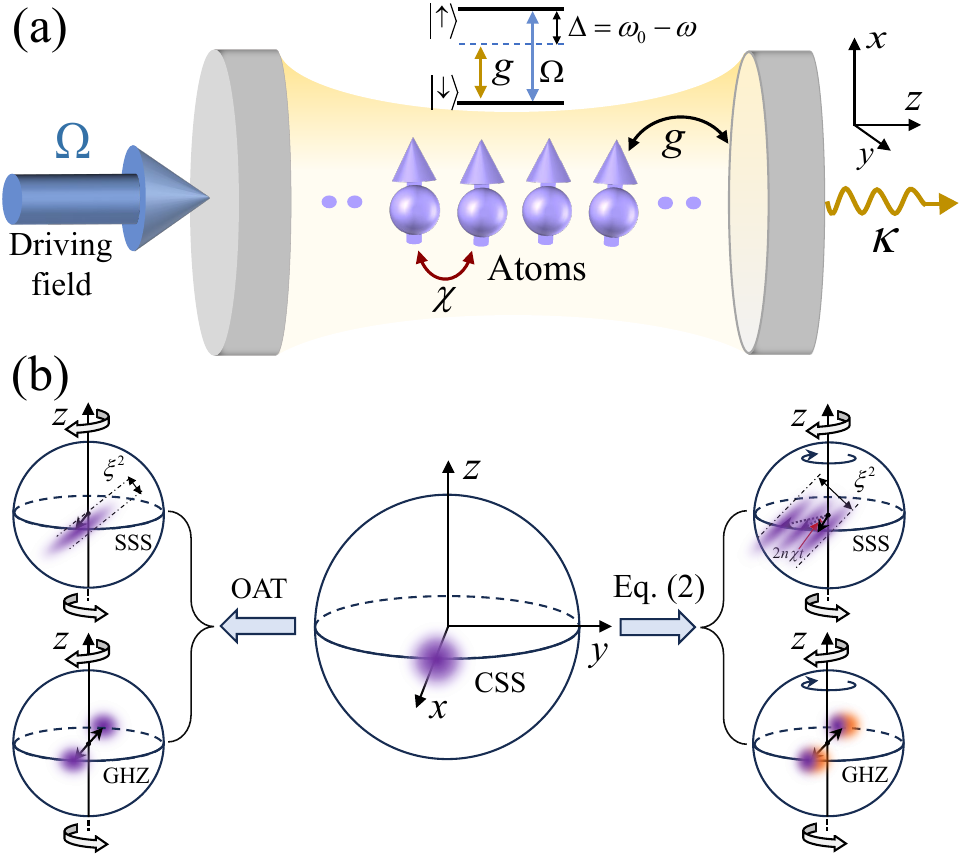}
	\caption{ (a) Schematic of the setup for realizing unitary OAT interaction. Atoms are confined in a cavity with linewidth $\kappa$, and the cavity mode $\hat a$ couples off-resonantly to each atom (with detuning $\Delta$), driving transitions from the ground state $\ket{\uparrow}$ to the excited state $\ket{\downarrow}$. The cavity system is driven by a classical control field $\Omega$. (b) The unitary OAT evolution generates a pure SSS for $\chi t \ll 1$ and a GHZ state at $\chi t = \pi/2$ (left column), whereas the nonunitary evolution results in mixed entangled states (right column). The Stark term induces dephasing of the OAT state, which broadens the spin uncertainty along the squeezed direction and results in the admixture of the ideal GHZ state with another orthogonal GHZ state (see main text).}
	\label{fig1}
\end{figure}

Although the Stark term can also be engineered to generate OAT dynamics, its inherent influence cannot be eliminated in the process, also leading to non-unitary OAT evolution \cite{schleier2010squeezing}. To eliminate the Stark term, Leroux \emph{et al}. proposed injecting specific quantum states (e.g., squeezed states) of the light field or employing a double-pass scheme, while Zhang \emph{et al}. suggested suppressing it by detuning the probe light far from the cavity resonance \cite{PhysRevA.91.033625}. Perhaps the most straightforward method is to employ a spin-echo scheme by directly applying $\pi$-pulses to atoms \cite{PhysRevLett.110.156402}, utilizing dynamical decoupling (DD) to disentangle the light and atoms, thereby restoring unitary OAT evolution. However, DD implemented via direct driving of the atomic ensemble faces challenges:
first, spatial inhomogeneity in the control field introduces additional noise that accumulates over the decoupling sequence, ultimately degrading the unitarity of the OAT dynamics; second, the inherent free Hamiltonian of the atoms can interfere with decoupling accuracy. To circumvent these issues, our work proposes a novel approach for eliminating the unwanted Stark term. We consider an ensemble of atoms coupling to a cavity mode via a standard TC interaction. Instead of driving the atoms directly, we apply a classical drive to the cavity mode (see Fig. \ref{fig1}). The cavity drive induces a collective precession of the ensemble about the $x$-axis, which, when exploited for spin control, enables the near-perfect suppression of the Stark term and thus the realization of near-unitary OAT dynamics. We demonstrate that both constant and time-varying drive can suppress the Stark term, with the latter yielding a more stronger OAT interaction. We also analyze the impact of atomic dissipation and reveal that the time-varying approach exhibits significantly greater resistance to decoherence.

The rest of the paper is organized as follows. In \cref{sec2}, we introduce the physical system and derive the effective Hamiltonian for atom-light interaction. We then examine in detail the influence of the Stark term on the generation of entangled states. In \cref{sec3}, we present both constant- and time-varying-driving approaches to realize near-unitary OAT evolution. In \cref{sec4}, we analyze the effects of atomic dissipation. Finally, a summary is provided in \cref{sec5}. The Appendix provides detailed derivations of the spin squeezing and effective interaction.

%\section{\label{sec:level2}Non-unitary OAT evolution}

\section{Nonunitary OAT evolution\label{sec2}}
The conventional approach for realizing OAT evolution employs the photon (or phonon) mode as a mediator to induce nonlinear interactions between spins. The most widely used spin-light interaction for engineering OAT dynamics is the TC interaction ($\hbar=1$) \cite{dong2022dynamics,fink2009dressed,PhysRev.170.379}
 \begin{table}[t]
	\renewcommand{\arraystretch}{1.2}
	\caption{Different bosonic input states and their corresponding  characteristic functions. Coherent state: $\alpha$ being the amplitude; Thermal state: $\bar{n}$ being the average photon (phonon) number; Squeezed state: $r$ being the squeezing parameter and $n$ being an even photon number.}
	\label{table1}
	\begin{ruledtabular}
		\begin{tabular}{lcc}
			\textrm{Input states} &
			\textrm{$|c_n|^2$} &
			\textrm{${\phi _{m,m'}}( t )$} \\
			\hline
			Fock state&$\delta_{n,n_0}$&${e^{ 2i{n_0}\chi t\left(m - m'\right)}}$\\
			\hline
			Coherent state&${e^{ - |\alpha {|^2}}}\frac{{{\alpha ^{2n}}}}{{n!}}$&${e^{ - |\alpha {|^2}\left[1 - {e^{ 2i\chi t(m - m')}}\right]}}$\\
			\hline
			Thermal state &$\frac{\bar{n}^n}{(1+\bar{n})^{n+1}}$&$\frac{1}{{1 + \bar n\left[1 - {e^{ 2i\chi t(m - m')}}\right]}}$\\
			\hline
			Squeezed state &$\frac{{{{(\tanh r)}^{n}}}}{{{2^{n}}\cosh r}}\frac{{n!}}{{{{[(n/2)!]}^2}}}$&$\frac{1}{{\sqrt {{{\cosh }^2}r - {{\sinh }^2}r{e^{ i4\chi t(m - m')}}} }}$\\
		\end{tabular}
	\end{ruledtabular}
\end{table}
\begin{eqnarray}
	\hat H_{{\rm{TC}} }=\omega \hat a^\dag \hat a+\omega_0 \hat S_z+g(\hat a^\dag \hat S_-+\hat a \hat S_+),\label{eq1}
\end{eqnarray}
where $\hat a$ ($\hat a^\dag$) denotes the annihilation (creation) operator of the bosonic mode with frequency $\omega$ and $g$ represents the coupling strength. $ \hat{S}_{\pm} = \hat{S}_x \pm i\hat{S}_y$ are the collective raising and lowering operators of spins, where $\hat{S}_k = \frac{1}{2} \sum_{j=1}^N \hat{\sigma}_k^{(j)} $ ($k \in \{x, y, z\}$), with \( \hat{\sigma}_k^{(j)} \) and $N$ being the standard Pauli operators acting on the \( j \)th spin and the number of spin (with a splitting frequency $\omega_0$), respectively. To clearly elucidate the core physics, let us for the moment neglect the decay of the cavity mode (that is, $\kappa=0$). In a rotating frame with respect to $\hat H_0=\omega \hat a^\dag \hat a+\omega_0 \hat S_z$, we have $\hat H_{\rm{TC}}^{\rm{rot}} =g{\hat S_ + }\hat a\exp (i\Delta t) + h.c.$ with the detuning $\Delta  = {\omega _0} - \omega $. In the case of $\Delta \gg g$, one obtains the effective Hamiltonian \cite{james2000quantum}
\begin{eqnarray}
    {\hat H^{\rm{eff}}_{\rm{TC}}} = -2\chi{\hat a^\dag }\hat a{\hat S_z} + \chi \hat S_z^2,\label{eq2}
\end{eqnarray}
with $\chi  = {g^2}/{\Delta }$, where the first term represents the AC-Stark shifts on atomic levels induced by the off-resonant coupling and the second quadratic term is the well-known OAT Hamiltonian. The interaction (\ref{eq2}) has been realized in various platforms \cite{braverman2019near,leroux2010implementation,leroux2012unitary,colombo2022time,schleier2010squeezing}.
We emphasize that the first Stark term alone can serve as a resource for generating OAT dynamics \cite{schleier2010squeezing,braverman2019near,leroux2010implementation,leroux2012unitary,li2022collective,schleier2010states}. By applying a monochromatic drive to an optical cavity coupled to an atomic ensemble via the ${\hat a^\dag }\hat a{\hat S_z}$ interaction \cite{schleier2010squeezing}, one can realize atoms-light dynamics similar to Eq. (\ref{eq2}) \cite{leroux2012unitary}. Nevertheless, the Stark term in Eq. (\ref{eq2}) is detrimental, as it induces dephasing in the atomic system and results in a nonunitary OAT evolution. Previous studies employing the TC approach to OAT engineering eliminate the influence of the first Stark term by assuming that the bosonic mode occupies a specific state, such as a Fock or squeezed state \cite{zheng2010waveguide,rivera2023creating}, or by claiming that dynamic decoupling can suppress it  \cite{rahman2024learning,zhao2014room,zhang2021generation,groszkowski2020heisenberg}. However, in reality, the bosonic state is typically a multi-photon (phonon) state, such as a coherent state. In such cases, the influence of the first AC-Stark term on OAT evolution becomes non-negligible and should be carefully estimated to avoid state-preparation errors.

To do so, we assume that the spin ensemble is initially prepared in the coherent spin state (CSS) $\ket{\frac{\pi}{2},0}=\sum_{m=-S/2}^{S/2}\mathcal{C}_m\ket{S,m}$ \cite{kitagawa1993squeezed}, where the coefficients $\mathcal{C}_m=\frac{1}{2^S}\binom{2S}{S+m}^\frac{1}{2}$ with $S=N/2$ being the total collective angular
momentum, while the bosonic mode is initially in an arbitrary supposition state $\ket{\psi_b}=\sum_n c_n\ket{n}$, where $\ket{n}$ denotes the Fock state of $\hat a$ and the coefficients $c_n$ are normalized such that $\sum_n |c_n|^2=1$. Given the input states and the interaction defined by Eq. (\ref{eq2}), one can derive the unitarily evolved state at time $t$,
\begin{eqnarray}
    \ket{\Psi(t)}&=&\exp({{-it\hat H^{\rm{eff}}_{\rm{TC}}}})\ket{\frac{\pi}{2},0} \ket{\psi_b}\nonumber\\
    &=&\sum\limits_{m,n} {\mathcal{C}_m{c_n}{e^{ - i\chi tm\left( {m -2 n} \right)}}\left| {S,m} \right\rangle \left| n \right\rangle }. \label{eq3}
\end{eqnarray}
Taking a partial trace over mode $\hat a$ results in
\begin{eqnarray}
{\hat\rho _S}(t) &=& \sum\limits_{m,m'} \mathcal{C}_m\mathcal{C}_{m'}^* {e^{ - i\chi t({m^2} - {{m'}^2})}} \nonumber\\
 &&\times{\phi _{m,m'}}\left( t \right)\left| {S,m} \right\rangle \left\langle {S,m'} \right|,
\label{eq4}
\end{eqnarray}
where we have defined the characteristic function (CF) ${\phi _{m,m'}}( t ) = \sum\nolimits_n {{{\left| {{c_n}} \right|}^2}{e^{  i2n\chi {\mathop{\rm t}\nolimits} (m - m')}}}$. Obviously, the influence of the Stark term on the OAT evolution is determined by the CF ${\phi _{m,m'}}( t )$. The CFs of different types of input bosonic states are given in Table \ref{table1}. For the case of a Fock state $\ket{n_0}$, the obtained spin state remains a pure state, $\ket{\Psi_S(t)}=\sum_{m}\mathcal{C}_me^{i2n_0\chi tm}e^{-i\chi tm^2}\ket{S,m}=e^{i2n_0\chi t \hat S_z}e^{-i\chi t \hat S_z^2}\ket{\pi/2,0}$, which represents a state that first experiences an OAT evolution and then is rotated around the
$z$-axis by an angle of $2n_0\chi t$. However, for the case of a coherent state $\ket{\alpha}$, the obtained spin state for small $\chi t(m-m')$ is approximately

\begin{eqnarray}
{{\hat \rho }_S}(t) &\approx& {\sum\limits_{m,m'} {{{{\cal C}_m}{\cal C}_{m'}^*} } }{e^{ - i\chi t({m^2} - {{m'}^2})}}{e^{ i2|\alpha {|^2}\chi t(m - m')}}\nonumber\\
 &&\times {e^{ - 4|\alpha {|^2}\chi^2 t^2{{(m - m')}^2}/2}}\left| {S,m} \right\rangle \langle S,m'|,\label{eq5}
\end{eqnarray}
from which one can see that, in addition to the rotation of the collective spin around the $z$-axis [induced by the factor ${e^{ i2|\alpha {|^2}\chi t(m - m')}}$], the OAT state will also undergo coherence dissipation at a rate proportional to the amplitude of the coherent state [induced by the factor ${e^{ - 4|\alpha {|^2}\chi^2t^2{{(m - m')}^2}/2}}$].
This is because the bosonic mode and the spins get entangled after the interaction, and the tracing operation on the light fields leads to the leakage of spin information, which in turn drives the spins into a mixed state. Such a process is very similar to the scenario---the interaction between the spins and a reservoir, which, as we know, will lead to the dissipative
dynamics of the spin subsystem. Therefore, the coherent-state input bears a \emph{nonunitary} OAT evolution. For other input states, such as the thermal and squeezed states, the CFs in the weak-coupling regime are proportional to $1/[1+4\bar n^2\chi^2t^2(m-m')^2]$ and $1/[\cosh r\sqrt{2\chi t (m'-m)}]$ ($r\gg 1$), respectively, which therefore will also cause the dissipation of the spin subsystem and lead to nonunitary OAT evolution.

\subsection{Spin squeezing}

To see explicitly how the dissipation impacts the OAT evolution, we first employ the generation of SSS through the interaction $\hat H^{\rm{eff}}_{\rm{TC}}$. Consider the special case that the input state of the bosonic mode is a coherent state, then the evolved spin state is described by the state in Eq. (\ref{eq5}). For simplicity, we neglect the linear rotation evolution in Eq. (\ref{eq5}) as it can be canceled by simply applying a homogeneous magnetic field along the $z$-axis. Then, from Eq. (\ref{eq5}) one may derive the expectation values of the collective spin at time $t$
\begin{eqnarray}
\left\langle \hat S_x(t)\right\rangle &=& e^{-\frac{\mu'}{2}} S\cos ^{2 S-1}\left(\frac{\mu}{2}\right),\\\label{eq6}
\left\langle\hat{S}_y^2(t)\right\rangle &=& \frac{S}{2}\left(S+\frac{1}{2}\right)\nonumber\\&&- e^{-2\mu'}\frac{S}{2}\left(S-\frac{1}{2}\right) \cos ^{2 S-2}(\mu),\\\label{eq7}
{\left\langle {{\hat T_{y,z}}} (t)\right\rangle} &=& -2{e^{ - \frac{\mu'}{2}}}S\left( {S - \frac{1}{2}} \right)\nonumber\\
 &&\times {\cos ^{2S - 2}}\left(\frac{\mu}{2}\right)\sin \left(\frac{\mu}{2}\right),\label{eq8}
\end{eqnarray}
where we have defined ${\hat T_{y,z}} = {\hat S_z}{\hat S_y} + {\hat S_y}\hat S_z$, $\mu=2\chi t$, and $\mu'=4|\alpha|^2\chi^2t^2$. These results indicate that the dissipation induced by the bosonic mode leads to the decay of not only the macroscopic spin $\langle \hat S_x\rangle$ but also the correlation term $\langle \hat T_{y,z}\rangle$, which equals zero for a CSS ($\chi t=0$) and serves as the original of spin squeezing \cite{kitagawa1993squeezed}. As a result, the dissipation process competes with OAT (squeezing) evolution, which, as we show below, leads to a reduction in the maximal achievable squeezing.

To quantify the spin squeezing, we use the squeezing parameter given by Kitagawa and Ueda \cite{kitagawa1993squeezed}: ${\xi^2} = 4{\text{min}(\Delta {\hat S_{\theta }})^2}/N$, where $\hat S_\theta=\hat S_y\cos\theta+\hat S_z\sin\theta$ denotes the collective spin operator along the $\theta$ direction in the $y$-$z$ plane \cite{ma2011quantum}. A SSS is created when $\xi^2<1$ ($\xi^2=1$ for a CSS).
To find out the minimal variances of $\hat S_{\theta}$, the expectation values $\langle {\hat S_\theta ^2} \rangle  = {(\Delta {{\hat S}_\theta })^2} = [\langle {\hat S_z^2} \rangle  + \langle {\hat S_y^2} \rangle  + \sqrt {(\langle {\hat S_y^2} \rangle  - \langle {\hat S_z^2} \rangle)^2  + \langle {\hat T_{y,z}} \rangle^2 } \cos 2(\theta  - \phi )]/2$ needs to be evaluated, where $\phi  = \arctan [(\langle {\hat S_y^2} \rangle  - \langle {\hat S_z^2}\rangle )/\langle {\hat T_{y,z}} \rangle ]/2$ and the first equality is valid provided that $\langle\hat S_z\rangle=\langle\hat S_y\rangle=0$ for the spin state of Eq. (\ref{eq5}), where the rotation effect is neglected for simplicity. The variance can be minimized when $\cos2(\theta-\phi)=-1$, resulting in
\begin{eqnarray}
{(\Delta {{\hat S}_{\rm{min}} })^2} = \frac{S}{2}\left[ {1 + \frac{1}{2}\left( {S - \frac{1}{2}} \right)\mathcal{A}\left( {1 - \sqrt {1 + \frac{{{\mathcal{B}^2}}}{{{\mathcal{A}^2}}}} } \right)} \right],\nonumber\\\label{eq9}
\end{eqnarray}
where we have defined $\mathcal{A} = 1 - {e^{ - 2\mu'}}{\cos ^{2S - 2}}(\mu)$ and
$\mathcal{B} = 4{e^{ - \frac{\mu'}{2}}}{\cos ^{2S - 2}}(\mu/2)\sin (\mu/2)$. For the case of large $S\gg 1$ and small $\mu\ll 1$, there will be $\mathcal{B}/\mathcal{A}\simeq 2\mu\exp [-(2\mu ' + S{\mu ^2})/4]\ll 1$. Then, one can Taylor expand the variances of Eq. (\ref{eq9}) around ${\mathcal{B}}/{\mathcal{A}}\approx 0$ to give
\begin{eqnarray}
{(\Delta {{\hat S}_{\rm{min}} })^2} &\simeq& \frac{S}{2}\left[ {1 + \frac{1}{2}\left( {S - \frac{1}{2}} \right)\left( {\frac{{{\mathcal{B}^4}}}{{8{\mathcal{A}^3}}} - \frac{{{\mathcal{B}^2}}}{{2\mathcal{A}}}} \right)} \right]\nonumber\\
 &\simeq& \frac{S}{2}\left[ {\frac{{\delta '}}{{\delta ' + {\delta ^2}}} + \frac{{{\delta ^4}}}{{{{4\left( {\delta ' + {\delta ^2}} \right)}^3}}}} \right.\left. { + \frac{2}{3}{\beta ^2}} \right],\label{eq10}
\end{eqnarray}
where we set $\delta  = S\mu /2,\delta ' = S\mu '/2$, and $\beta  = S{\mu ^2}/4$. For a vacuum input ($\delta'=0$), Eq. (\ref{eq10}) is reduced to ${(\Delta {{\hat S}_{\rm{min}} })^2} \simeq S[{1}/{({4{\delta ^2}})} + {2}{\beta ^2}/{3}]/2$, which is exactly the minimal variance created by the unitary OAT evolution and can lead to a maximal squeezing $\xi_{\rm{OAT}}^2\propto S^{-2/3}$  \cite{kitagawa1993squeezed,ma2011quantum}. For a coherent-state input, if we choose $\delta\sim\delta'\gg 1$, the second term of Eq. (\ref{eq10}) is negligible, resulting in the amount of spin squeezing $\xi^2=4{(\Delta {{\hat S}_{{\rm{min}}}})^2}/N \simeq 1/\delta  + 2{\beta ^2}/3$, which can be minimized to give $\xi^2 \simeq 5/(4\sqrt[5]{{12}}){S^{-2/5}}$ obtained at $\mu_{\rm{min}}=12^{1/5}S^{-3/5}$.
Evidently, the presence of the Stark term has led to a worse spin-squeezing scaling, due to the broadening of the uncertainty along the squeezing direction.

Such an uncertainty broadening can also be understood qualitatively in the following way. As schematically shown in Fig. \ref{fig1}(b), when a bosonic superposition state is used as the input, the quantum state described by Eq. (\ref{eq3}) can be rewritten as  $\left| {\Psi (\mu)} \right\rangle  = \sum\nolimits_n {{c_n}\left| {{\rm{OAT}_\mu},n\mu } \right\rangle \left| n \right\rangle }$, where $\left| {{\rm{OAT}_\mu},\phi }  \right\rangle  = e^{ i\phi \hat S_z}e^{ - i\mu \hat S_z^2}\left| {\frac{\pi} {2},0} \right\rangle $ denotes an OAT state whose mean-spin direction (MSD) is rotated by an angle $\phi$ around the $z$-axis from the $x$-axis direction \cite{ma2011quantum}. This implies that the Stark term will cause a rotation of the OAT state around the $z$ axis. Depending on the Fock number $n$, the OAT state rotates different angle. The larger the value of $n$, the greater the rotation angle of the OAT state. Correspondingly, the reduced density operator of Eq. (\ref{eq4}) can be reexpressed as
\begin{eqnarray}
{\hat \rho _S}(\mu) = \sum\limits_{n=0}^\infty {{{\left| {{c_n}} \right|}^2}\left| {{\rm{OAT}_\mu},n\mu } \right\rangle \left\langle {{\rm{OAT}_\mu},n\mu } \right|},\label{eq11}
\end{eqnarray}
 which reveals that there exists a MSD uncertainty as if the OAT state ``spreads out'' along the equator of the Bloch sphere---a phenomenon closely analogous to the dephasing process inherent in spin echo dynamics \cite{PhysRev.80.580}. It is precisely the MSD uncertainty that degrades the spin squeezing. This uncertainty is evidently determined by the distribution of the probability amplitude $|c_n|^2$ of the bosonic input state.
\subsection{GHZ-state generation}

Another important application of the OAT dynamics is to create the GHZ state \cite{kitagawa1993squeezed}. Next, we will consider the influence of the Stark term to the GHZ-state generation. It is well known that the GHZ state can be created if we apply the OAT evolution to the CSS
$\ket{\pi/2,0}$ for a time duration $t=\pi/2\chi$, that is, $\ket{\rm{GHZ}}=\left| {{\rm{OAT}_\pi},0} \right\rangle=\left| {\frac{\pi}{2},\Phi } \right\rangle  - i\left| {\frac{\pi}{2},\Phi+\pi} \right\rangle$ with $\Phi=-N\pi/2$ \cite{PhysRevA.56.2249}. Analogous to the spin squeezing, the Stark term can also induce the dephasing of the GHZ state, resulting in the distortion of the target state, as depicted in Fig. \ref{fig1}(b). For $\mu=\pi$, Eq. (\ref{eq11}) becomes
\begin{eqnarray}
{{\hat \rho }_S}(\pi ) &=& \sum\limits_{n = 0}^\infty  {\left( {{{\left| {{c_{2n}}} \right|}^2}\left| {{\rm{GHZ}}} \right\rangle \langle {\rm{GHZ}}|} \right.} \nonumber\\
&&\left. { + {{\left| {{c_{2n + 1}}} \right|}^2}{e^{ - i\pi {{\hat S}_z}}}\left| {{\rm{GHZ}}} \right\rangle \langle {\rm{GHZ}}|{e^{i\pi {{\hat S}_z}}}} \right).
\end{eqnarray}
 In addition to the GHZ state generated by the unitary OAT evolution, this state is also mixed with another type of GHZ state---namely, a new type of GHZ state created by rotating the original OAT-generated GHZ state by $\pi$ around the $z$-axis. Since these two types of states are orthogonal to each other, the purity of the target state $\ket{\rm{GHZ}}$ is thereby greatly reduced.
To characterize the performance of GHZ-state preparation, we utilize the fidelity $\mathcal{F}=\bra{\psi_T}\hat \rho\ket{\psi_T}$, which measures how close the state $\hat\rho$ is to the state $\ket{\psi_T}$. Here $\ket{\psi_T}=\ket{\rm{GHZ}}$ and $\hat\rho$ is given by Eq. (\ref{eq11}), then one can calculate the fidelity, and get getting
\begin{eqnarray}
\mathcal{F} &=& \sum\limits_{n = 0}^\infty  {{{\left| {{c_n}} \right|}^2}{{\left| {\left\langle {{\rm{GHZ}}} \right.\left| {{\rm{OA}}{{\rm{T}}_\pi },n\pi } \right\rangle } \right|}^2}} \nonumber\\
 &=& \sum\limits_{n = 0}^\infty  {{{\left| {{c_n}} \right|}^2}{{\cos }^{4S}}\left( {\frac{{n\pi }}{2}} \right)} \nonumber\\
 &=& \sum\limits_{n = 0}^\infty  {{{\left| {{c_{2n}}} \right|}^2}},
\end{eqnarray}
which for a coherent-state input is of the form
\begin{eqnarray}
\mathcal{F} = \frac{1}{2} + \frac{1}{2}{e^{ -2 {{\left| \alpha  \right|}^2}}},
\end{eqnarray}
showing that the fidelity of the GHZ state created by the interaction (\ref{eq2}) decreases exponentially with $|\alpha|$.

The above analysis reveals that for a general bosonic superposition state, the Stark term has a negative impact on both spin squeezing and GHZ-state preparation, leading to a nonunitary OAT evolution. It is therefore essential to develop simple and efficient methods to suppress the Stark term and achieve unitary OAT dynamics.

\section{NEAR-Unitary OAT evolution\label{sec3}}
One approach to suppressing the Stark term involves extracavity control: for example, injecting non-classical states of light (e.g., squeezed states) into the atom-cavity system  \cite{schleier2010squeezing}; employing a lossless-delay-line-based double-pass scheme \cite{schleier2010squeezing}; or utilizing far-off-resonance driving \cite{PhysRevA.91.033625}. An alternative is to apply spin-echo control, such as a sequence of global $\pi$-pulses, to dynamically decouple the Stark term \cite{PhysRevLett.110.156402}. Obviously, methods relying on non-classical resources impose relatively high demands on experimental conditions, whereas the far-off-resonance technique attenuates the effective OAT interaction strength. The spin-echo approach, on the other hand, demands extremely precise pulse control; imperfections like spatial inhomogeneity or intensity fluctuations in $\pi$-pulses can introduce deleterious noise.

In this section, we propose a hybrid scheme that combines the merits of both strategies. By driving the cavity mode with a classical optical field, we induce a linear rotational evolution in the atomic ensemble, which is harnessed to achieve DD of the Stark term. Our scheme operates entirely with classical resources. Crucially, because both the requisite linear rotation and the desired nonlinear OAT evolution are generated in parallel via the same cavity mode, it inherently avoids the noise associated with direct coherent pulse manipulation of atoms. This makes the protocol suitable for a wide range of bosonic-atom quantum systems.
 \begin{figure}[t]
	\centering
	\includegraphics[width=0.98\linewidth]{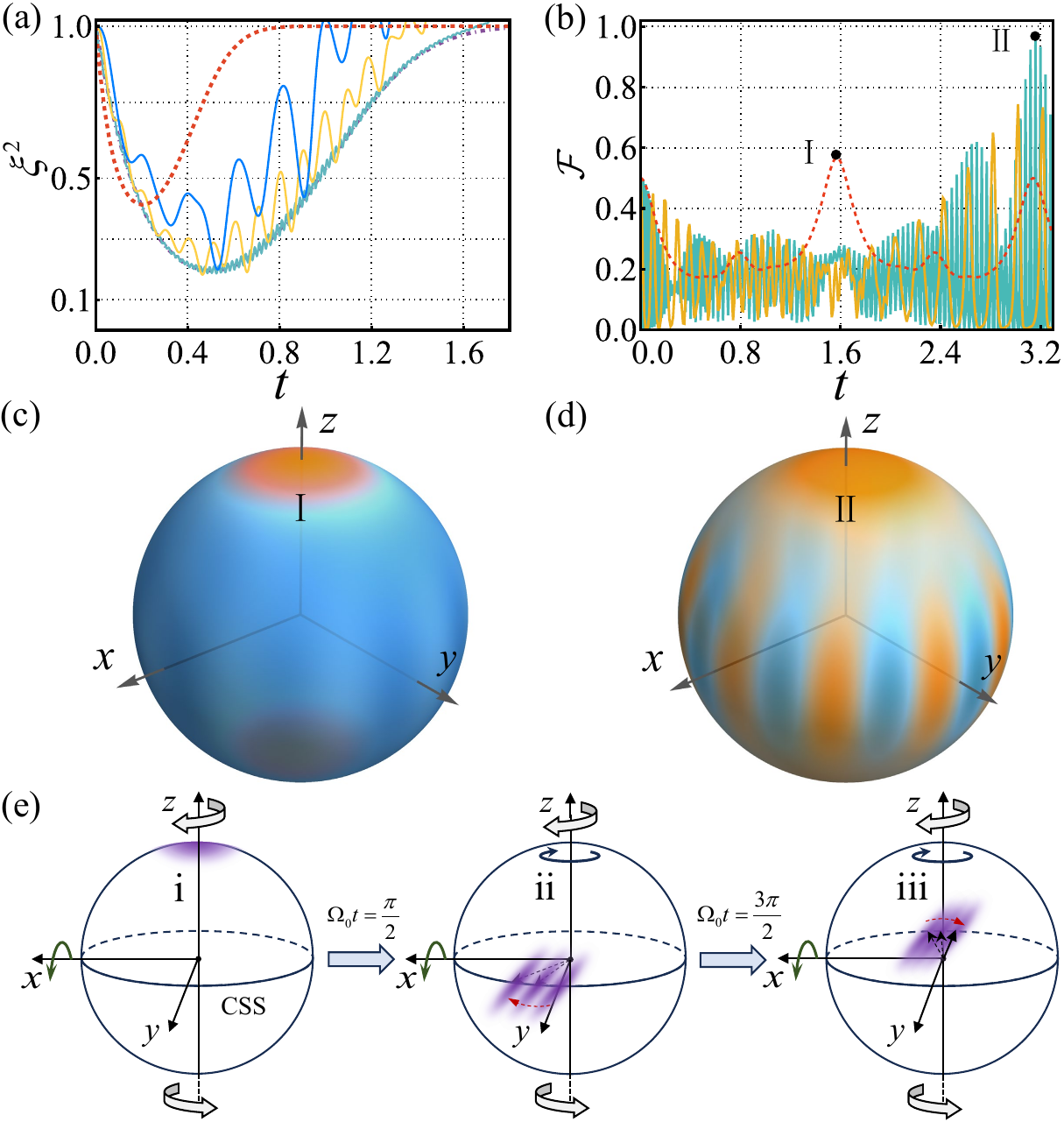}
	\caption{ (a) Time evolution of the spin-squeezing parameter $\xi^2$ induced by the ideal OAT Hamiltonian (purple dotdashed line),  the Hamiltonian of Eq. (\ref{eq2}) (red dashed line), and the Hamiltonian of Eq. (\ref{eq16}) for constant driving amplitudes $\Omega_0=16$ (blue solid line), $\Omega_0=32$ (orange solid line), and  $\Omega_0=160$ (cyan solid line). (b) Time evolution of the fidelity of GHZ-state preparation induced by the Hamiltonian of Eq. (\ref{eq2}) (red dashed line) and the Hamiltonian of Eq. (\ref{eq16}) for constant driving amplitudes $\Omega_0=32$ (orange solid line) and $\Omega_0=160$ (cyan solid line). Plot of the Winger functions for the peak states I (c) and II (d) in (b). (e) Schematic of unitary OAT evolution under a constant drive. The initial CSS has its MSD aligned along the $z$-axis.
The drive induces rapid Larmor precession of the MSD about the $x$-axis. Dephasing of the OAT state occurs when the MSD points along $+y$, and rephasing occurs when it points along $-y$. The net effect cancels the Stark term, leaving the system in a pure OAT evolution. Here we take $N=10$, and the bosonic mode is in a coherent state with $\alpha=1$, and $t$ is the unit of $1/\chi$ while $\Omega_0$ is in unit of $\chi$.}
	\label{fig2}
\end{figure}

Next, we will show how to implement near-unitary OAT evolution by utilizing a driven TC model. To simplify the discussion and highlight the core physics, we focus on a specific physical system.
Consider a system of two-level atoms coupled to a single cavity mode via the TC interaction described above, as shown in Fig. \ref{fig1}(a). In addition, the cavity mode is driven by an external laser field at frequency $\omega_d$. As a result, the atom-field interaction can be described by
\begin{eqnarray}
{\hat H_{{\rm{DTC}}}} = {\hat H_{{\rm{TC}}}} + \Omega(t) \left(\hat a{e^{i{\omega _d}t}} + {\rm{h}}{\rm{.c}}{\rm{.}}\right),\label{eq14}
\end{eqnarray}
where $\Omega(t)$ denotes the time-varying driving amplitude.
Moving to a rotating frame with respect to $\omega_d(\hat S_z+\hat a^\dag \hat a)$, we obtain
\begin{eqnarray}
\hat H_{{\rm{DTC}}}^{{\rm{rot}}} &=& \Delta' {{\hat a}^\dag }\hat a + g\left( {{{\hat a}^\dag }{{\hat S}_ - } + \hat a{{\hat S}_ + }} \right) \nonumber\\&&+ \Omega(t) \left( {\hat a + {{\hat a}^\dag }} \right),\label{eq15}
\end{eqnarray}
where $\Delta'  = \omega  - {\omega _d}$ and we have assumed that the driving field is resonant with the atomic transition $\omega_d = {\omega _0}$. Under the condition of large detuning $\Delta'\gg g$, one can approximately diagonalize  $\hat H_{{\rm{DTC}}}^{{\rm{rot}}}$ by using  a Schrieffer-Wolff transformation \cite{PhysRevLett.110.156402,PhysRev.149.491} $\hat H_{{\rm{DTC}}}^{{\rm{eff}}} = {e^{\hat R}}\hat H_{{\rm{DTC}}}^{{\rm{rot}}}{e^{ - \hat R}}\simeq \hat H_{{\rm{DTC}}}^{{\rm{rot}}} + [\hat R,\hat H_{{\rm{DTC}}}^{{\rm{rot}}}] + [\hat R,[\hat R,\hat H_{{\rm{DTC}}}^{{\rm{rot}}}]]/2$ with $\hat R=g( {{{\hat a}^\dag }{{\hat S}_ - } - \hat a{{\hat S}_ + }} )$, resulting in the effective Hamiltonian
\begin{eqnarray}
\hat H_{{\rm{DTC}}}^{{\rm{eff}}} = {\tilde\Omega}(t){\hat S_x} - 2\chi {\hat a^\dag }\hat a{\hat S_z} + \chi \hat S_z^2,\label{eq16}
\end{eqnarray}
where $\tilde\Omega(t)=2g\Omega(t)/\Delta'$. Compared to the interaction (\ref{eq2}), an additional linear term, $\tilde \Omega (t){\hat S_x}$, now appears. This term will induce a rotation of the atomic system around the $x$-axis, which, as we shall show later, can be used to suppress the spin dephasing caused by the Stark term. We point out here that although the interaction described by Eq. (\ref{eq16}) can be realized by directly applying an oscillating field to the atoms, such an approach may induce additional quantum noise such as phase diffusion \cite{zhou2009quantum} (see Appendix B for more details).

\subsection{Constant driving}
Let us first consider the case of a constant driving, such that $\tilde\Omega(t)= \Omega_0$. In the rotating frame
defined by $\hat V_1(t)=\exp{(-i\Omega_0 t\hat S_x)}$, the Hamiltonian (\ref{eq16}) becomes
\begin{eqnarray}
\hat{\tilde{H}}_{{\rm{DTC}}}^{{\rm{eff}}}  &=& \hat V_1\left( t \right)\hat H_{{\rm{DTC}}}^{{\rm{eff}}}{{\hat V}^\dag_1 }\left( t \right) - i\hat V_1\left( t \right){\partial _t}{{\hat V}^\dag_1 }\left( t \right)\nonumber\\
 &=& \frac{\chi}{2}\left( {\hat S_z^2 + \hat S_y^2} \right) - 2\chi {{\hat a}^\dag }\hat a\left[ {{{\hat S}_z}\cos ({\Omega _0}t)} \right.\nonumber\\
&&\left. { - {{\hat S}_y}\sin ({\Omega _0}t)} \right] + \frac{\chi }{2}\left[ {\left(\hat S_z^2-\hat S_y^2\right)\cos (2{\Omega _0}t)} \right.\nonumber\\
&&\left. { - \{ {{\hat S}_z},{{\hat S}_y}\}\sin (2{\Omega _0}t)} \right]\nonumber\\
&=& \frac{\chi}{2}\left[ {\hat S_z^2 + \hat S_y^2}  + \sum\limits_{m = 1}^2 {\left( {{{\hat h}_m}{e^{im{\Omega _0}t}} + \text{h.a.}} \right)}\right],
\end{eqnarray}
where $\{.\}$ denotes the anti-commutator product and we have defined the new atomic operators
\begin{eqnarray}
{{\hat h}_1} &=&  - 2{{\hat a}^\dag }\hat a\left( {{{\hat S}_z} + i{{\hat S}_y}} \right),\nonumber\\
{{\hat h}_2} &=& \frac{1}{2}\left[ {\hat S_z^2 -\hat S_y^2+i {\{ {{\hat S}_z},{{\hat S}_y}\} }} \right].\nonumber
\end{eqnarray}
We are interested in a time-averaged dynamic over a period $T$ that is significantly longer than the spin rotation period around the $x$-axis, i.e., $T\gg 2\pi/\Omega_0$. Then, one obtains the effective Hamiltonian \cite{james2000quantum}
\begin{eqnarray}
\hat{\tilde{H}}_{{\rm{DTC}}}^{{\rm{eff}}} &\simeq & \frac{\chi}{2}\left( {\hat S_z^2 + \hat S_y^2} \right) + \sum\limits_{m = 1}^2 {\frac{1}{{m{\Omega _0}}}\left[ {\hat h_m^\dag ,{{\hat h}_m}} \right]} \nonumber\\
 &=& \frac{\chi}{2}\left( {\hat S_z^2 + \hat S_y^2} \right) + \frac{ 2{\chi ^2}}{{{\Omega _0}}}{\left( {{{\hat a}^\dag }\hat a} \right)^2}{{\hat S}_x}.\label{eq18}
\end{eqnarray}
If $\Omega_0\gg \chi S$, the last term in Eq. (\ref{eq18}) are negligible. As a consequence, we arrive at
\begin{eqnarray}
\hat{\tilde{H}}_{{\rm{DTC}}}^{{\rm{eff}}} \simeq \frac{\chi}{2}\left( {\hat S_z^2 + \hat S_y^2} \right) =-\frac{\chi}{2}\hat S_x^2,\label{eq19}
\end{eqnarray}
where in the second equality we have used the relation $\hat S_x^2+\hat S_y^2+\hat S_z^2=S(S+1)$ and disregarded the constant term. Obviously, the interaction (\ref{eq19}) is precisely the OAT interaction. Therefore, the constant driving of the TC model can eliminate the Stark term and thus yield near-unitary OAT evolution, but at the expense of reducing the coupling strength by a factor of two. Using the previously established physical picture, the cancellation of the Stark term can be qualitatively interpreted as follows. Instead, the initial spin state is now prepared in the CSS $\ket{0,0}$, as shown in Fig. \ref{fig2}(e)(i). The driving term induces a rapid rotation of the CSS around the $x$-axis. During this rotation, the Stark term in Hamiltonian (\ref{eq16}) causes the dephasing of the atoms, as illustrated in Fig. \ref{fig2}(e)(ii). However, due to the opposite dephasing (or rephasing \cite{PhysRev.80.580}) along the negative $y$-axis, the cumulative effect of the Stark-induced atomic dephasing cancels out over multiple evolution cycles [see Fig. \ref{fig2}(e)(iii)].
In Fig. \ref{fig2}(a) and (b) we compare the evolution results under the approximation Hamiltonian of Eq. (\ref{eq19}) with the results obtained from Eq. (\ref{eq16}). In the case of large $\Omega_0$, both the squeezing and fidelity curves confirm that the evolution induced by Eq. (\ref{eq16}) is equivalent to a unitary OAT evolution. Notably, for small values of $\Omega_0$, the degree of squeezing can even exceed that induced by OAT. This phenomenon arises because the nonlinear evolution generated in this regime takes a form intermediate between OAT and two-axis countertwisting spin squeezing \cite{liu2011spin}.

\subsection{Time-varying driving}
\begin{figure}[t]
	\centering
	\includegraphics[width=0.99\linewidth]{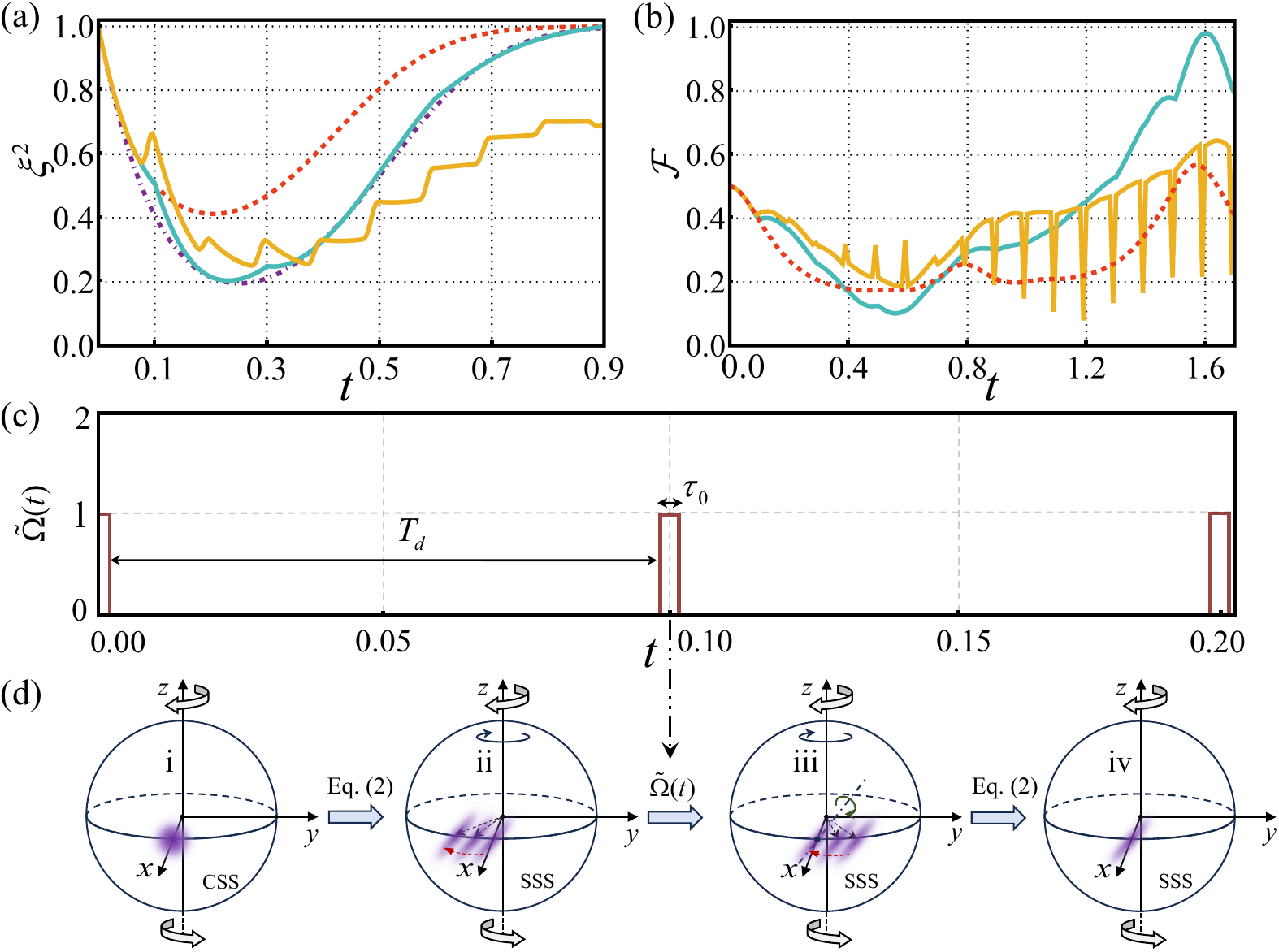}
	\caption{ (a) Time evolution of the spin-squeezing parameter $\xi^2$ induced by the ideal OAT Hamiltonian (purple dotdashed line),  the Hamiltonian of Eq. (\ref{eq2}) (red dashed line), and the Hamiltonian of Eq. (\ref{eq16}) for time-varying driving $d=0.3$ (orange solid line), $0.01$ (cyan solid line). (b) Time evolution of the fidelity of GHZ-state preparation induced by time-varying driving with $d=0.2$ (orange solid line), $0.04$ (cyan solid line), $0$ (red dashed line). (c) The amplitude distribution of the driving field in the time domain. (d) Schematic of unitary OAT evolution under a time-varing drive. The initial CSS has its MSD aligned along the $x$-axis. The time-varying drive induces Larmor precession of the MSD about the $x$-axis. In the absence of the driving field, the OAT state undergoes dephasing. The narrow rectangular pulses instantaneously rotate the collective spin by an angle $\pi$ around the $x$-axis, after which rephasing begins. The net effect cancels the Stark term, leaving the system in a purely OAT-evolved state. Here we take $N=10$, $T_d=0.1$, and $\Omega_0=0$. The bosonic mode is in a coherent state with $\alpha=1$ and $t$ is in unit of $1/\chi$.}
	\label{fig3}
\end{figure}
For the above scheme, the MSD (that is, $z$-axis) of the atomic ensemble is always perpendicular to the rotation axis (that is, $x$-axis). However, in a realistic implementation, keeping the MSD parallel to the rotation axis would be more advantageous. For example, it is less susceptible to decoherence \cite{PhysRevLett.133.173604} and can maintain the spin squeezing for a longer time \cite{PhysRevA.63.055601}. To realize such configuration, we assume that the driving field is a time-modulated field with an amplitude of the form
\begin{eqnarray}
{\tilde\Omega}\left( t \right) = {\Omega _0} + {{\cal A}_0}d\left[ {1 + 2\sum\limits_{n =1}^\infty  {{\rm{sinc}}\left( {\pi nd} \right)\cos ({n\omega t})} } \right],\nonumber
\end{eqnarray}
where $\omega$ denotes the modulation frequency and ${\cal A}_0$ refers to the periodical modulation amplitude. In the time domain, the second term corresponds to square waves that have a width $\tau_0$ and are $T_d=\pi/\omega$ away from each other, as shown in Fig. \ref{fig3}(c). $d=\tau_0/T_d$ denotes the duty cycle.
In a rotating frame with
\begin{eqnarray}
\hat V_2(t) &=& T{e^{i\int_0^t {{{\cal A}_0}d\left[ {1 + 2\sum\nolimits_{n = 1}^\infty  {{\rm{sinc}}\left( {\pi nd} \right)\cos \left( {n\omega t} \right)} } \right]\hat S_xd\tau } }}\nonumber\\
 &=& {e^{i\left[ {{{\cal A}_0}dt + \frac{{2{{\cal A}_0}}}{{\pi \omega }}\sum\nolimits_{n = 1}^\infty  {\frac{{{\rm{sin}}\left( {\pi nd} \right)\sin \left( {n\omega t} \right)}}{{{n^2}}}} } \right]\hat S_x}},
\end{eqnarray}
where ${\cal T}$ denotes the time-ordering operator, the Hamiltonian becomes
\begin{eqnarray}
\hat{\tilde{H}}_{{\rm{DTC}}}^{{\rm{eff}}} &=& {\Omega _0}\hat S_x + \frac{\chi }{2}\left( {\hat S_z^2 + \hat S_y^2} \right)\nonumber\\
 &&- 2\chi {{\hat a}^\dag }\hat a\left[ {{{\hat S}_z}\cos ({{\tilde \Omega }_0}t)} \right.\left. { - {{\hat S}_y}\sin ({{\tilde \Omega }_0}t)} \right]\nonumber\\
 &&+ \frac{\chi }{2}\left[ {\left( {\hat S_z^2 - \hat S_y^2} \right)\cos (2{{\tilde \Omega }_0}t)} \right.\left. { - \{ {{\hat S}_z},{{\hat S}_y}\} \sin (2{{\tilde \Omega }_0}t)} \right],\nonumber\\\label{eq21}
 \end{eqnarray}
where we have defined the new parameter ${{\tilde \Omega }_0} ={A_0}d + 2{A_0}\sum\nolimits_{n = 1}^\infty {\rm{sin}}( {n\pi d} )\sin ( {n\omega t} )/({n^2}\pi \omega t)$. Using the Poisson summation formula
\begin{eqnarray}
\sum\limits_{n = 1}^\infty  {\frac{{\cos \left( {nx} \right)}}{{{n^2}}}}  = \frac{{{\pi ^2}}}{6} - \frac{{\pi x}}{2} + \frac{{{x^2}}}{4},\quad 0 \le x \le 2\pi,\nonumber
\end{eqnarray}
one obtains
\begin{eqnarray}
{{\tilde \Omega }_0}t = {{\cal A}_0}dt + \frac{{2{{\cal A}_0}}}{{\pi \omega }}{\cal S}(d,\omega t\bmod 2\pi ),\label{eq22}
\end{eqnarray}
where
\begin{eqnarray}
{\cal S}(d,\theta)=\begin{cases}
\dfrac{\pi\theta}{2}(1-d), & 0\le\theta\le\pi d,\nonumber\\[6pt]
\dfrac{\pi d}{2}(\pi-\theta), & \pi d<\theta\le\pi(2-d).
\end{cases}
\end{eqnarray}
Consider the case that $d\ll 1$, such that ${\cal S}(d,\theta ) \approx \pi d(\pi  - \theta )/2$. As a result, Eq. (\ref{eq22}) at time $t=mT+\tau$ ($m=0,1,2...$ and $0\leq\tau< T$) is found to be
\begin{eqnarray}
{{\tilde \Omega }_0}t &=& m{{\cal A}_0}dT + {{\cal A}_0}d\tau  + \frac{{2{{\cal A}_0}}}{{\pi \omega }}\frac{{\pi d}}{2}(\pi  - \omega \tau )\nonumber\\
 &=& \frac{{{{\cal A}_0}d}}{\omega }\left( {2m + 1} \right)\pi,
\end{eqnarray}
which is independent of $\tau$, implying that ${{\tilde \Omega }_0}t$ behaves as a step function. If ${{{{\cal A}_0}d}}/{\omega }=1/2$, then ${{\tilde \Omega }_0}t=({m + 1/2})\pi$. Substituting it into Eq. (\ref{eq21}), we find
\begin{eqnarray}
\hat {\tilde H}_{{\rm{DTC}}}^{{\rm{eff}}} &=& {\Omega _0}{{\hat S}_x} + \frac{\chi }{2}\left( {\hat S_z^2 + \hat S_y^2} \right)\nonumber\\
 &&+ 2\chi {{\hat a}^\dag }\hat a{{\hat S}_y}\cos (m\pi ) - \frac{\chi }{2}\left( {\hat S_z^2 - \hat S_y^2} \right)\cos (2m\pi )\nonumber\\
 &=& {\Omega _0}{{\hat S}_x} + \chi \hat S_y^2 + {\left( { - 1} \right)^m}2\chi {{\hat a}^\dag }\hat a{{\hat S}_y}\nonumber\\
 &\simeq& {\Omega _0}{{\hat S}_x} + \chi \hat S_y^2,\label{eq24}
\end{eqnarray}
where in the last equality we assumed $\omega$ is large and therefore neglected the last rapidly oscillating Stark term. Equation (\ref{eq24}) is exactly the twist-and-turn interaction, which has been shown to create multiparticle entanglement
much more rapidly than the OAT interaction \cite{PhysRevA.92.023603,PhysRevA.97.053618,PhysRevA.105.062456}. It is worth noting that the time-varying driving keeps the coupling strength of the OAT interaction unchanged. The mechanism for canceling the Stark term is illustrated schematically in Fig. \ref{fig3}(d). The initial spin state is prepared in the CSS $\ket{\frac{\pi}{2},0}$, as shown in Fig. \ref{fig3}(d)(i). During the inactive part of the duty cycle, the Stark term causes dephasing of the atomic state, as illustrated in Fig. \ref{fig3}(d)(ii). The active part instantaneously rotates the collective spin by an angle $\pi$ around the $x$-axis [see Fig. \ref{fig3}(d)(iii)], after which the spin state begins to rephase. By periodically repeating this sequence, the OAT state is always confined without spreading out, thereby suppressing the dynamics induced by the Stark term [see Fig. \ref{fig3}(d)(iv)].

In Fig. \ref{fig3}(a) and (b) we compare the evolution results under the approximation Hamiltonian of Eq. (\ref{eq24}) with the results obtained from Eq. (\ref{eq16}) for $\Omega_0=0$. In the case of small $d$, both the squeezing and fidelity curves confirm that the evolution induced by Eq. (\ref{eq16}) is equivalent to a unitary OAT evolution.
\begin{figure}[t]
	\centering
	\includegraphics[width=0.98\linewidth]{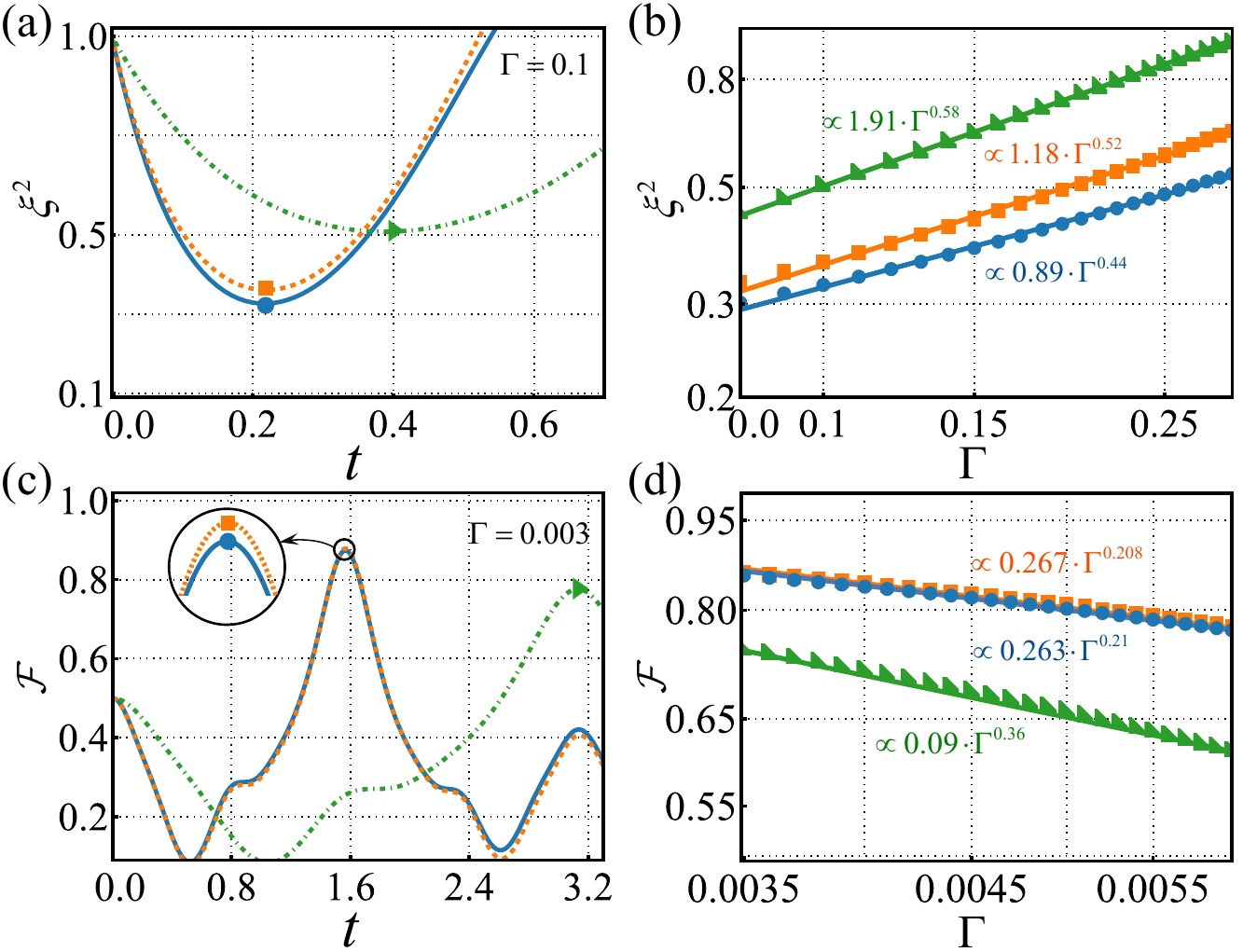}
	\caption{ The spin squeezing (a) and the GHZ-state fidelity (c) versus time $t$ in the presence of dissipations calculated from the master equations (\ref{eq26}) (green dotdashed line) and (\ref{eq27}) (blue solid line). The orange dashed line is also calculated from Eq. (\ref{eq26}), but with the coupling strength increased by a factor of two.  The maximal achievable squeezing (b) and the fidelities (d) as a function of dissipation $\Gamma$. Here we take $N=10$ and $t$ is in unit of $1/\chi$ while $\Gamma$ is in unit of $\chi$.}
	\label{fig4}
\end{figure}

\section{Dissipation analysis\label{sec4}}

In reality, it is inevitable that the bosonic mode undergoes dissipations, as illustrated in Fig. \ref{fig1}(a). We assume the bosonic excitations leak with a rate $\kappa$, described by the Lindbladian $\kappa{\cal L}(\hat a,\hat \rho )$, where $ {\cal L}(\hat A,\hat \rho )= \hat A\hat \rho {\hat A^\dag } - \{ {\hat A^\dag }\hat A,\hat \rho \} /2$. Starting from Eq. (\ref{eq15}) and taking the limit $\Delta'\gg\kappa\gg g$, one may adiabatically eliminate the bosonic mode and obtain the spin-only master equation \cite{pineiro2022emergent,sundar2024squeezing}
\begin{eqnarray}
\frac{d}{{dt}}{{\hat \rho }_S} =  - i\left[ {\hat H_{{\rm{DTC}}}^{{\rm{eff}}},{{\hat \rho }_S}} \right] + \Gamma\mathcal{L}( {{{\hat S}_ - },{{\hat \rho }_S}} ),\label{eq25}
\end{eqnarray}
where $\Gamma=\chi\kappa/\Delta'$. In addition to the bosonic-mode-mediated collective
interactions, the bosonic dissipation also induces the collective decay of atoms. For the case of constant drive, in a rotating frame at the drive frequency $\Omega_0$ by defining $\hat{\tilde{\rho}}_S=\hat V_1\hat\rho\hat V_1^\dag$, the master equation becomes
\begin{eqnarray}
\frac{d}{{dt}}{{\hat {\tilde\rho} }_S} &=& i\left[ {\frac{\chi }{2}\hat S_x^2,{{\hat {\tilde\rho} }_S}} \right] + \Gamma {\cal L}( {{{\hat S}_x},{{\hat {\tilde\rho} }_S}} )\nonumber\\
 &&+ \frac{\Gamma }{2}{\cal L}( {{{\hat S}_y},{{\hat {\tilde\rho} }_S}} ) + \frac{\Gamma }{2}{\cal L}( {{{\hat S}_z},{{\hat {\tilde\rho} }_S}} ),\label{eq26}
\end{eqnarray}
where in the Lindbladian part we have also neglected the oscillating terms. This result indicates that under strong driving, the original $\hat S_-$ dissipation is transformed into a triaxial decoherence process: with full decay rate along the $x$-direction, and half decay rate along the $y$- and $z$-directions. For the case of time-varying driving, in a rotating frame given by the transformation $\hat{\bar{\rho}}_S=\hat V_2\hat\rho\hat V_2^\dag$, the master equation is of the form ($\Omega_0=0$)
\begin{eqnarray}
\frac{d}{{dt}}{{\hat {\bar\rho} }_S} = -i\left[ {\chi\hat S_y^2,{{\hat {\bar{\rho}}}_S}} \right]
 + \frac{\Gamma }{2}{\cal L}( {{{\hat S}_-},{{\hat {\bar\rho}}_S}} ) + \frac{\Gamma }{2}{\cal L}( {{{\hat S}_+},{{\hat {\bar\rho}}_S}} ),\nonumber\\\label{eq27}
\end{eqnarray}
showing that the periodic driving also induces the collective $\hat S_+$ dissipation. Apparently, the two decoupling schemes lead to fundamentally different types of dissipation. In Fig. \ref{fig4}, we compare the performance of Eqs. (\ref{eq26}) and (\ref{eq27}) and show that the time-varying scheme is significantly more efficient. We also compare the above two types of dissipative mechanisms and find that the dissipative mechanism of Eq. (\ref{eq27}) can better preserve the spin squeezing, while the one in Eq. (\ref{eq26}) shows an advantage in the preparation of GHZ states.

\section{Conclusion\label{sec5}}

In summary, this work revisits the TC-model-based approach for generating OAT dynamics in an atomic ensemble. We provide a detailed analysis of the detrimental effects caused by the intrinsic Stark term associated with this method, particularly on the generation of atomic entangled states, including SSSs and GHZ states. Our findings reveal that when the bosonic mode is initialized in a coherent state, the Stark term significantly degrades spin-squeezing performance and causes an exponential decay in the fidelity of GHZ-state generation. These adverse effects underscore the importance of suppressing the Stark term. To address this issue, we propose an alternative strategy using a tailored classical driving field applied to the atoms-cavity hybrid system. We demonstrate that the proposed approach enables near-unitary OAT dynamics: the constant-driving scheme reduces the OAT coupling strength by half, while the time-varying-driving scheme preserves the full nonlinear interaction strength. Furthermore, we examine the robustness of the scheme against atomic dissipation and find that, compared to the constant drive, the time-varying drive exhibits stronger resistance to noises. Our scheme exploits the decoupling effect arising from fast spin rotation induced by cavity driving. Unlike conventional spin-echo methods that address dephasing accumulated over time  \cite{PhysRev.80.580} or rely on direct atomic driving \cite{PhysRevLett.110.156402}, our approach allows for rapid and accurate decoupling of the light-atom entanglement. Consequently, the loss of atomic information may remain modest even in the presence of cavity dissipation. Therefore, the proposed DD schemes is expected to perform well in cavity-atoms systems with cavity driving and dissipation \cite{schleier2010squeezing}. A detailed investigation of its performance in dissipatively driven regimes is left to future studies.
We expect that the proposed scheme will offer useful insights for future research in atomic-ensemble-based quantum information processing and quantum precision measurement.

\begin{acknowledgments}
This work was supported by the Innovation Program for Quantum Science and Technology (2023ZD0300904) and the Natural Science Foundation of Zhejiang province, China (Grant No.LMS25A040004).
\end{acknowledgments}

%\bibliography{REF}

	\appendix
	\section{Derivation of the spin squeezing}
In this Appendix, we present the detailed derivation of Eq. (\ref{eq10}). To further simply the minimal variance
	\begin{eqnarray}
					{(\Delta {{\hat S}_{\rm{min}} })^2} \simeq \frac{S}{2}\left[ {1 + \frac{1}{2}\left( {S - \frac{1}{2}} \right)\left( {\frac{{{\mathcal{B}^4}}}{{8{\mathcal{A}^3}}} - \frac{{{\mathcal{B}^2}}}{{2\mathcal{A}}}} \right)} \right],\label{A1}
	\end{eqnarray}
it is necessary to evaluate the quantities $\mathcal{A}$ and $\mathcal{B}$. Consider the case $S\gg 1$, then one approximately has
\[\mathcal{A} \simeq 1 - {e^{ - \vartheta {\mu ^2}}},\mathcal{B}\simeq 2\mu {e^{ - \vartheta {\mu ^2/4}}},\]
where we have defined $\vartheta  = 2|\alpha {|^2} + S - 1$. Assuming that $\vartheta\mu^2\ll 1$ and $\mu\ll 1$, the ratios in Eq. (\ref{A1}) can be approximated as
\begin{eqnarray}
\frac{{{\mathcal{B}^2}}}{{2\mathcal{A}}} &=& \frac{{2{\mu ^2}{e^{ - \vartheta {\mu ^2}/2}}}}{{1 - {e^{ - \vartheta {\mu ^2}}}}} \simeq \frac{2}{\vartheta } - \frac{{\vartheta {\mu ^4}}}{{12}},\\
\frac{{{\mathcal{B}^4}}}{{8{\mathcal{A}^3}}} &=& \frac{{2{\mu ^4}{e^{ - \vartheta {\mu ^2}}}}}{{{{\left( {1 - {e^{ - \vartheta {\mu ^2}}}} \right)}^3}}} \simeq \frac{1}{{{\vartheta ^2}}} + \frac{2}{{{\vartheta ^3}{\mu ^2}}},
\end{eqnarray}
where we kept the terms up to the second order in $\vartheta\mu^2$.
Substituting them into Eq. (\ref{A1}) yields
\begin{eqnarray}
{(\Delta {{\hat S}_{\rm{min}} })^2} \simeq \frac{S}{2}\left( {1 - \frac{S}{\vartheta } + \frac{S}{{2{\vartheta ^2}}} + \frac{S}{{{\vartheta ^3}{\mu ^2}}} + \frac{\vartheta {\mu ^4}S}{{24}}} \right).\label{A4}
\end{eqnarray}
It is reasonable to assume that $S\gg |\alpha|^2$, then one can neglect the third term in Eq. (\ref{A4}) as it is proportional to $1/S$, resulting in
\begin{eqnarray}
{(\Delta {{\hat S}_{\min }})^2} &\simeq& \frac{S}{2}\left[ {\frac{{2|\alpha {|^2}}}{{2|\alpha {|^2} + S}} + \frac{S}{{{{\left( {2|\alpha {|^2} + S} \right)}^3}{\mu ^2}}} + \frac{{S^2}{\mu ^4}}{{24}}} \right]\nonumber\\
 &\simeq& \frac{S}{2}\left[ {\frac{{\delta '}}{{\delta ' + {\delta ^2}}} + \frac{{{\delta ^4}}}{{4{{\left( {\delta ' + {\delta ^2}} \right)}^3}}} + \frac{2}{3}{\beta ^2}} \right],
\end{eqnarray}
which is the result given in Eq. (\ref{eq10}) in the main text.

\section{Dynamical decoupling via direct atomic driving}
In this Appendix, we present calculations of the detrimental effects on DD arising from the direct application of the driving field to the atoms. We consider the driving TC model
\begin{eqnarray}
\hat H = \omega {\hat a^\dag }\hat a + {\omega _0}{\hat S_z} + g\left( {{{\hat a}^\dag }{{\hat S}_ - } + \hat a{{\hat S}_ + }} \right)+ \Omega (t){\hat S_x},\label{B1}
\end{eqnarray}
where $\Omega (t)$ denotes the time-varying driving amplitude of $\hat S_x$. After transforming to the interaction picture, the Hamiltonian in Eq. (\ref{B1}) becomes
\begin{eqnarray}		
		{{\hat{\tilde{H}}}} &=& g({{\hat a}^\dag }{{\hat S}_ - }{e^{i\Delta t}} + \hat a{{\hat S}_ + }{e^{ - i\Delta t}})\nonumber\\
		&+& \frac{\Omega (t) }{2}({{\hat S}_ + }{e^{i{\omega _0}t}} + {{\hat S}_ - }{e^{ - i{\omega _0}t}}), \label{B2}
\end{eqnarray}
where \(\Delta  = \omega  - {\omega _0}\). Considering the constant driving \(\Omega (t) = {\Omega _0}\), we obtain the effective Hamiltonian
\begin{eqnarray}
       \hat{H}_{\rm{eff}} =  - 2\chi {{\hat a}^\dag }\hat a{{\hat S}_z} + \chi \hat S_z^2 - \frac{{\Omega _0^2}}{{2{\omega _0}}}{{\hat S}_z}. \label{B3}
\end{eqnarray}
This result indicates that directly applying a constant $S_x$ driving to the TC model yields an effective $S_z$ evolution rather than the desired $S_x$ rotation. To realize linear $\hat S_x$ dynamics, one may subject the atoms to a rapidly oscillating field, $\Omega (t) = {\Omega _0}\cos ({\omega _0}t)$, to atoms, yielding
\begin{eqnarray}
	    {\hat {\tilde{H}}} &=&  g({{\hat a}^\dag }{{\hat S}_ - }{e^{i\Delta t}} + \hat a{{\hat S}_ + }{e^{ - i\Delta t}})\nonumber\\
		&+& \frac{{{\Omega _0}\cos ({\omega _0}t)}}{2} ({{\hat S}_ + }{e^{i{\omega _0}t}} + {{\hat S}_ - }{e^{ - i{\omega _0}t}}), \label{B4}
\end{eqnarray}
which leads to the effective Hamiltonian
\begin{eqnarray}
    \hat H_{\rm{eff}} =  - 2\chi {{\hat a}^\dag }\hat a{{\hat S}_z} + \chi \hat S_z^2 + \frac{{{\Omega _0}}}{2}{{\hat S}_x} - \frac{{\Omega _0^2}}{{16{\omega _0}}}{{\hat S}_z}. \label{B5}	
\end{eqnarray}
Although the desired linear ${{\hat S}_x}$ dynamics is successfully produced, it simultaneously introduces an unwanted additional term proportional to ${{\hat S}_z}$, which must be canceled. It is important to note that applying a rapidly oscillating drive may itself adversely affect the atomic system \cite{dooley2016hybrid, li2013motional}. On one hand, in regions of avoided crossings, the system can undergo non-adiabatic transitions, leading to population leakage from the initial state into other levels \cite{shevchenko2010landau, silveri2017quantum}. On the other hand, such driving can amplify atomic decoherence and enhance phase diffusion \cite{zhou2009quantum}.

\end{document}